\documentclass[sigconf, authorversion=true]{acmart}

\copyrightyear{2021}
\acmYear{2021}
\setcopyright{acmlicensed}
\acmConference[EdgeSys'21]{4th International Workshop on Edge Systems, Analytics and Networking}{April 26, 2021}{Online, United Kingdom}
\acmBooktitle{4th International Workshop on Edge Systems, Analytics and Networking (EdgeSys'21), April 26, 2021, Online, United Kingdom}
\acmPrice{15.00}
\acmDOI{10.1145/3434770.3459733}
\acmISBN{978-1-4503-8291-5/21/04}

\usepackage{svg}
\usepackage{microtype}
\usepackage{csquotes}
\usepackage{ifthen}
\usepackage{hyperref}
\usepackage{subcaption}

\usepackage[margin=10pt,font=small,labelfont=bf]{caption}

\begin{document}

\title{Detecting and Mitigating Network Packet Overloads on Real-Time Devices in IoT Systems}
\date{\today}

\author{Robert Danicki}
\authornote{Both authors contributed equally to this research.}
\email{r.danicki@tu-berlin.de}
\affiliation{%
  \institution{Technische Universität Berlin}
  \streetaddress{Straße des 17. Juni 135}
  \city{Berlin}
  \country{Germany}
  \postcode{10623}
}

\author{Martin Haug}
\authornotemark[1]
\email{m.haug@tu-berlin.de}
\orcid{0000-0003-3136-8724}
\affiliation{%
  \institution{Technische Universität Berlin}
  \streetaddress{Straße des 17. Juni 135}
  \city{Berlin}
  \country{Germany}
  \postcode{10623}
}

\author{Ilja Behnke}
\email{i.behnke@tu-berlin.de}
\orcid{0000-0002-2437-8994}
\affiliation{%
  \institution{Technische Universität Berlin}
  \streetaddress{Straße des 17. Juni 135}
  \city{Berlin}
  \country{Germany}
  \postcode{10623}
}

\author{Laurenz Mädje}
\email{maedje@campus.tu-berlin.de}
\affiliation{%
  \institution{Technische Universität Berlin}
  \streetaddress{Straße des 17. Juni 135}
  \city{Berlin}
  \country{Germany}
  \postcode{10623}
}

\author{Lauritz Thamsen}
\orcid{0000-0003-3755-1503}
\email{lauritz.thamsen@tu-berlin.de}
\affiliation{%
  \institution{Technische Universität Berlin}
  \streetaddress{Straße des 17. Juni 135}
  \city{Berlin}
  \country{Germany}
  \postcode{10623}
}

\begin{abstract}
Manufacturing, automotive, and aerospace environments use embedded systems for control and automation and need to fulfill strict real-time guarantees.
To facilitate more efficient business processes and remote control, such devices are being connected to IP networks.
Due to the difficulty in predicting network packets and the interrelated workloads of interrupt handlers and drivers, devices controlling time critical processes stand under the risk of missing process deadlines when under high network loads. Additionally, devices at the edge of large networks and the internet are subject to a high risk of load spikes and network packet overloads. 

In this paper, we investigate strategies to detect network packet overloads in real-time and present four approaches to adaptively mitigate local deadline misses.
In addition to two strategies mitigating network bursts with and without hysteresis, we present and discuss two novel mitigation algorithms, called Budget and Queue Mitigation.
In an experimental evaluation, all algorithms showed mitigating effects, with the Queue Mitigation strategy enabling most packet processing while preventing lateness of critical tasks.  
\end{abstract}

\maketitle

\section{Introduction}
Safety-critical application areas like plant floors, automotive and aerospace use embedded real-time systems for control, monitoring and automation \cite{hanninenPresentFutureRequirements2006, sharpChallengesSolutionsEmbedded2010}. With the advent of the industrial Internet of Things (IoT), industrial control systems are connected to large IP networks \cite{mirian2016internet}.
Receiving network traffic is accomplished by triggering interrupts for incoming network packets which preempts the process currently running on the assigned core independent of its priority.
However, the timing and load of network packets is difficult to predict. Connecting the used microcontrollers with low processing power to IP networks might expose critical infrastructure to the traffic patterns of a larger network or the internet. Hence, the impact of networking might break real-time guarantees. 

When a device is subject to high packet loads, local processes can be fully preempted by interrupt service routines (ISRs) and driver processes, amounting to a denial of service (DoS) whether maliciously or by network fault. In real-time scenarios this becomes especially relevant as unpredictable processing delays can have deadline breaking effects \cite{behnke2020}.
Mission critical and hard real-time devices must therefore detect high network loads and mitigate their consequences with limited processing resources.

%Network stacks in embedded networking devices often come bundled with closed source device drivers. As network driver code in most cases is non-modifiable, solutions to mitigate network-generated interrupt floods are non-trivial.
Mitigating the effect of network-generated interrupts from software in a real-time operating system is challenging as the timing impact of mitigation techniques themselves has to be kept minimal. They are furthermore restricted to react after interrupts have already occurred as this is controlled by hardware.

Addressing this, our paper presents:
\begin{itemize}
    \item Three metrics for detecting amounts of network traffic that may jeopardize the local real-time guarantees: The network interrupt count, receive queue fill state, as well as the lateness of critical processes.
    \item Four techniques to mitigate the impact of high packet loads while maintaining the network services on a best-effort basis. \emph{Burst Mitigation} caps packets received per time slice, \emph{Hysteresis Mitigation} puts lower and upper boundaries on earliness, the novel \emph{Budget Mitigation} calculates how much time is left for handling interrupts, and finally, the novel \emph{Queue Mitigation} uses the queue fill-state as an indicator for interrupt activity.
    \item Experiments evaluating the four mitigation techniques on the real-time operating system FreeRTOS.
\end{itemize}

\subsubsection*{Outline} Section \ref{sec:approach} presents three detection metrics and four mitigation techniques.
We then evaluate the approaches in Section \ref{sec:results}. The results are discussed in Section \ref{sec:evaluation}.
We give an overview of related work in Section \ref{sec:related-work}, while Section \ref{sec:conclusion} concludes this paper.

\section{Approach}
In the following section, we present detection techniques and mitigation algorithms to handle network-generated interrupt floods in real-time systems.
\label{sec:approach}
\subsection{Detection}
To prevent a critical task from being drowned by network interrupts, we first need to detect such a situation. There are several different direct or indirect metrics we can use to do that.

\subsubsection*{Early- or Lateness} The most direct metric is the critical task's early- or lateness.
In many real-time systems, a process periodically performs a critical computation, targeting to finish it within a fixed duration.
When the critical task completes this computation in time (i.e. in less time than the target duration), we have positive earliness, defined as the target duration minus the actual time spent.
When, however, the critical task exceeds its target duration, it incurs lateness, defined as how much longer it took than targeted. We will express lateness as a ratio of the critical task's target duration, e. g. $100\%$ lateness means it took twice as long as intended.

This metric very directly corresponds to what we are trying to detect but also has a few drawbacks: 
For once, it introduces latency as processes can only report earliness/lateness once per task cycle.
Thus, mitigation techniques might need to be overly cautious (disabling networking while there is still some earliness) because otherwise, it will react too late, when lateness already crept in.
Additionally, there is the more practical concern that the metric may be hard to come by in real systems since somehow, the metric must be reported. We may thus call mitigation techniques relying on this \emph{cooperative}.

\subsubsection*{Network Interrupt Count} A less direct metric is the number of incoming interrupts, discretized by dividing time into fixed time slices and counting network interrupts in them.

The number of interrupts can be easily counted since Interrupt Service Routines (ISRs) often run custom code anyway.
Since many network interrupts occur per time slice, the metric's resolution is equally high. Timing precision is not crucial because misattributing the first few packets to the passed time slice does not introduce significant errors.

The drawback of this metric is that it only correlates with the situations we are trying to prevent if certain preconditions are true:
The network interrupt count shows approximately how much strain the interrupts are putting on a CPU.
We can thus estimate overall system resource usage if the resource requirements of the critical task stay approximately the same for each of its cycles.
The first precondition, therefore, is that the critical task has to require the same amount of resources at all times --- mitigation techniques relying on this can not react elastically to load change.
Furthermore, this metric assumes that the processing of each packet takes about the same time.
Mitigation techniques using this metric can only be effective if the ensemble of incoming packets is homogeneous enough such that the assumption that each packet takes approximately the same time to process is either true or practical because of the regular distribution of packet response time.

\subsubsection*{Network Receive Queue Fill State} \label{sec:queue-fill-metric}
A third possible metric is detecting if the queue of received packets to be processed by the network driver is full or not.
The queue fill state shows whether the network driver can keep pace with the incoming packets.
If more packets arrive than the network driver can handle, the queue will fill up until it reaches maximum capacity; it would empty in the opposite case.

Of course, the queue's capacity impacts the quality of this metric.
With a queue capacity that is too low, the metric will show saturation even for short bursts of packets that are not representative of the overall traffic pace.
Low queue sizes will also lead to saturation if the scheduler did not wake up the network driver for some time.
If, conversely, the queue is too large it acts as a cushion and will delay alerting by filling up, allowing the network routines to stay activated for longer than is prudent.

The metric scales well in terms of packet response time deviation and elastic critical loads because it directly mirrors the ability of the network driver to process incoming traffic.
It has to be coupled with a mechanism like scheduler priority that moderates network driver execution such that an appropriate resource ceiling is found for the tasks.

\subsection{Mitigation Techniques}
\label{sec:mitigations}
In the following, we present four different mitigation techniques.
With the \emph{Burst Mitigation}, we suggest a technique that prevents unresponsiveness in case of short packet bursts.
The \emph{Hysteresis Mitigation} controls the networking tasks based on the early- or lateness of the critical task.
This idea taken further, the novel \emph{Budget Mitigation} attempts to balance out the networking part and the critical task by assigning a networking budget to the driver that is derived from the critical task's earliness.
Finally, the novel \emph{Queue Mitigation} takes a step back and explores a different way to detect high load, enabling a simple yet effective mitigation technique.

\subsubsection*{Burst Mitigation}
\begin{figure}
    \centering
    %\includesvg[width=\columnwidth]{explainers/burst.svg}
    \includegraphics[width=\columnwidth]{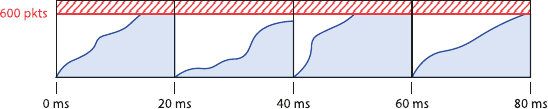}
    \caption{Burst Mitigation: Visualization of time slices.}
    \label{fig:burst}
\end{figure}

Burst Mitigation is a simple approach to deal with very high packet loads in short periods.
Conceptually, time is split into fixed time slices, each having a fixed maximum packet capacity.
The network ISR tracks these time slices by querying the operating system for the tick count each time it is invoked, starting a new slice if enough time has passed since the start of the last one.
In each slice, the number of packets is counted.
If this packet count surpasses the fixed capacity, network interrupts are disabled until the start of the next time slice.
We visualize this in Figure \ref{fig:burst}: The length of each time slice is $20ms$ and the curve plots the number of received packets in each slice.
The red zone on top is off-limits --- when $600$ packets are reached, interrupts are disabled.

Technically, the maximum number of packets that are accepted in a very short amount of time may be up to twice as large as the capacity limit.
To see this, consider the following case: When \emph{capacity}-many packets arrive directly at the end of one slice, the counter is reset immediately after that burst, allowing another burst of up to \emph{capacity}-many packets.
However, since the limit must be tuned manually anyway, this is not important conceptually.

\subsubsection*{Hysteresis Mitigation}
When the network load exceeds the trigger capacity for the Burst Mitigation by a small amount, network interrupt activation will oscillate quite a bit.
We used the number of packets per time frame to detect whether we had to act to maintain the local real-time guarantees.
This, however, is just an imperfect proxy for the metric we are actually interested in: The critical task's performance, i. e. the lateness of the critical task. Therefore, we based this mitigation on it.

Hysteresis is a popular technique for controlling processes \cite{krasnoselskiiStaticHysteron1989}.
It defines two thresholds: A maximum allowable threshold for a metric above which whatever action contributing to the rise of that metric will be ceased and a minimum threshold below which the action will be re-started.
We use the lateness/earliness metric reported by the critical task, i.e. how much time is left in the time slice when the task has been accomplished as the hysteresis control metric.
Once the earliness falls below the minimum allowable value, we stop processing new packets in the network driver, deactivate interrupts from there and wait (in a loop that sleeps for some time in every iteration) for the critical task to report an earliness higher than the threshold.

\begin{figure}
    \centering
    %\includesvg[width=.8\columnwidth]{explainers/hysteresis.svg}
    \includegraphics[width=.8\columnwidth]{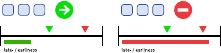}
    \caption{Hysteresis Mitigation: Late- / earliness on a scale with the block and unblock thresholds.}
    \label{fig:hysteresis-explain}
\end{figure}

\subsubsection*{Budget Mitigation}
This novel mitigation technique ties the earliness metric more closely to the amount of work that is permissible within the network subsystems.
For example, there could be a local critical task load that uses up $90\%$ of computation time per cycle.
Even a minuscule network load could then lead to the Hysteresis Mitigation permanently disabling the network.
Furthermore, we wanted to use the late-/earliness metric to react more responsively to elastic loads.

Budget Mitigation is another cooperative approach.
The critical task reports its earliness to the network driver after each completed cycle of computation.
This earliness is interpreted as the time budget of the network driver.
The driver will measure the time of its operations and subtract that from the latest budget.
Once the budget is depleted, the network subsystems will be suspended (including the interrupts) until a new earliness notification is issued.
For this mitigation to work, we set both the critical and the network driver task to equal priorities such that the network driver has a chance to deplete its budget after which it will actively yield to the critical task.
This is an important tweak since the Budget Mitigation acts as a specialized scheduler by deciding when the network subsystem has to cease to operate.
It would otherwise be defeated by the scheduler's time slice logic.

\begin{figure}
    \centering
    %\includesvg[width=\columnwidth]{explainers/budget.svg}
    \includegraphics[width=\columnwidth]{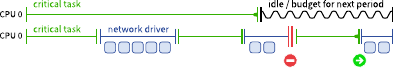}
    \caption{Budget Mitigation: Timelines of tasks on CPU. The network driver can preempt the critical task until its budget is depleted.}
    \label{fig:budget-explain}
\end{figure}

\subsubsection*{Queue Mitigation}
We looked at three mitigation techniques so far ---~a very simple approach that requires a manual definition of a capacity limit and two cooperative ones that require communication with the critical task.
With Queue Mitigation, we propose a new, more universal, yet effective non-cooperative mitigation. 

This mitigation technique is based on the simple observation with regards to the network queue we made in Section \ref{sec:queue-fill-metric}:
The network driver can keep up with the traffic when the packet queue is not full.
Thus, with queue mitigation, we simply disable network interrupts if they fail to put a packet into the queue (because it was already full) and only re-enable the interrupts from the driver once it has processed all queued packets.

When using this mitigation technique, we assign a higher priority to the critical task.
As a result, the network driver is the first process running out of time when CPU resources get scarce or the interrupt frequency rises.
The direct effect is that the queue fills up.
With a reasonable queue size, the networking is then disabled before the critical task is too stressed, thus protecting the critical task from interrupt overload.

\begin{figure}
    \centering
    %\includesvg[width=\columnwidth]{explainers/queue.svg}
    \includegraphics[width=\columnwidth]{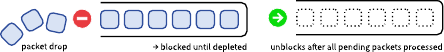}
    \caption{Queue Mitigation: After the packet queue fills up, all incoming packets are dropped until it is empty again.}
    \label{fig:queue-explain}
\end{figure}

\section{Evaluation}
\label{sec:results}

To investigate the effectiveness of the presented metrics and techniques, we test on the ESP32 microcontroller. The controller is a common ARM-based development board with two CPU cores and the lightweight FreeRTOS operating system. FreeRTOS provides basic task scheduler and interrupt management as well as some data structures for our application.
The multi-core platform allows us to separate the traffic generator from the traffic consumer by placing each on one of the two cores.
A local task simulating time-critical computation is additionally placed on the consuming core such that both compete for CPU time.

In our test setup, the interrupt handler fills up a FreeRTOS queue while the network driver empties it, thus processing the incoming data.
This allows us to analyze the effect of both, the interrupts and the driver on the critical computation load.

\begin{figure}[b]
    \centering
    %\includesvg[width=.8\columnwidth]{system.svg}
    \includegraphics[width=.8\columnwidth]{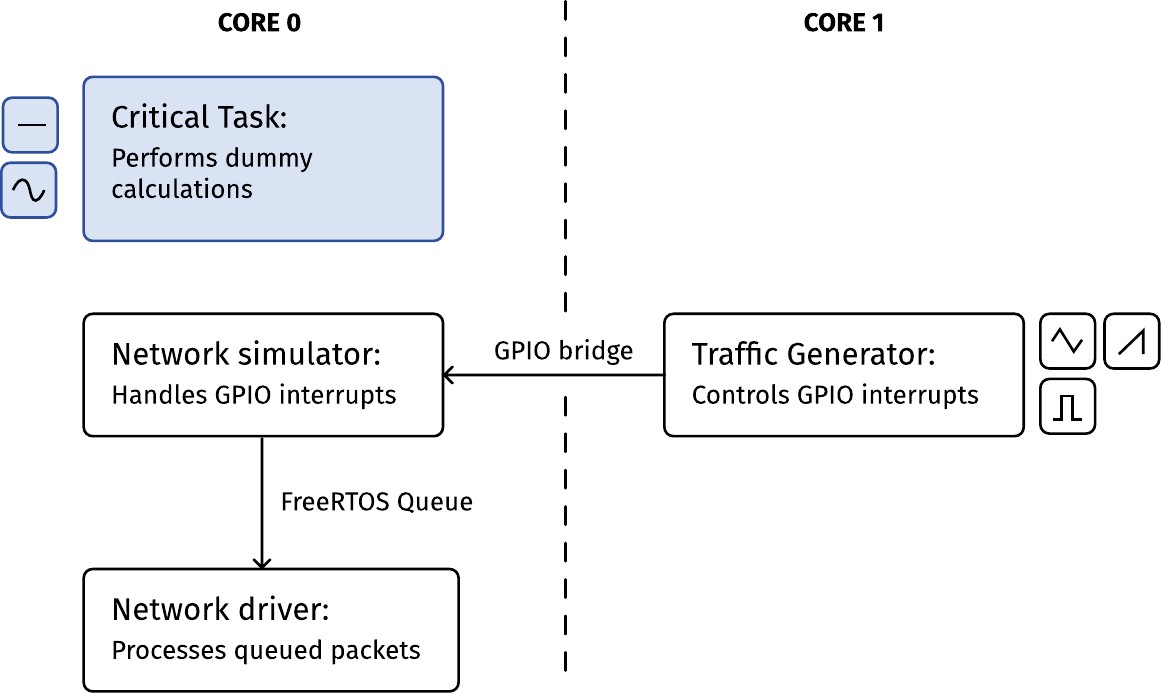}
    \vspace{0.5em}
    \caption{Setup used to simulate different interrupt load scenarios on the ESP32 SoC.}
    \label{fig:overview}
\end{figure}

% The traffic generator running on the second core can generate network loads of three load patterns (\emph{Sawtooth}, \emph{Pyramid}, \emph{Square}).

The traffic generator running on the second core generates network loads with a pyramid-shaped load pattern.
It sets a GPIO pin to high for each received packet, triggering a GPIO interrupt on the other core over a GPIO bridge.
There, the network simulator synthesizes a TCP SYN packet and places it into a queue for consumption by the network driver which acknowledges the packet's arrival.
Meanwhile, the observed critical task (also running on the first core) calculates an ascending series of binomial coefficients to generate an equal work load for each task cycle.
Its goal is to reach a target $(n, k)$ in a small time frame of $10ms$.
If the target is reached before the permissible time expired, the critical task will sleep the remaining time and start over in the next period.
If, however, the critical task fails to reach its target in time, it will accumulate lateness, i.e. continue until it reached the target values and then start its new period immediately. Figure~\ref{fig:overview} provides a graphical overview of this setup.

The amount of
\begin{itemize}
    \item interrupts triggered on the second core
    \item interrupts executed on the first core
    \item packets processed in the network driver
    \item critical task cycles
    \item accumulated critical task lateness
\end{itemize}
are saved into atomic variables of a monitoring routine running on core 1.
Every second, the controller reports these variables via the serial interface and then clears the counters.
%Our evaluation is based on this data that was collected by a computer connected to the controller via its serial interface.

The design imposes a few limitations: First, it does not account for interrupt sources other than network interrupts. However, in high network load situations, network interrupts should significantly outnumber other interrupts such that this is not too much of a problem (for reference, in a typical FreeRTOS system there are \~1000 background interrupts per second to control the scheduler, in contrast to \~100,000 network interrupts in high-load scenarios \cite{goyette2007analysis}).

Secondly, the critical task may only use one core since we need one of the ESP32's cores for traffic generation and analysis. We believe that this is not a significant drawback since most current microcontrollers remain single-cored \cite{bucaioniTechnologyPreservingTransitionSingleCore2017}.

\begin{figure}[t]
    \centering
    \begin{subfigure}[t]{0.48\columnwidth}
    %\includesvg[width=\textwidth]{plot_NONE_n400k350_CRITPRI_packets.svg}
    \includegraphics[width=\textwidth]{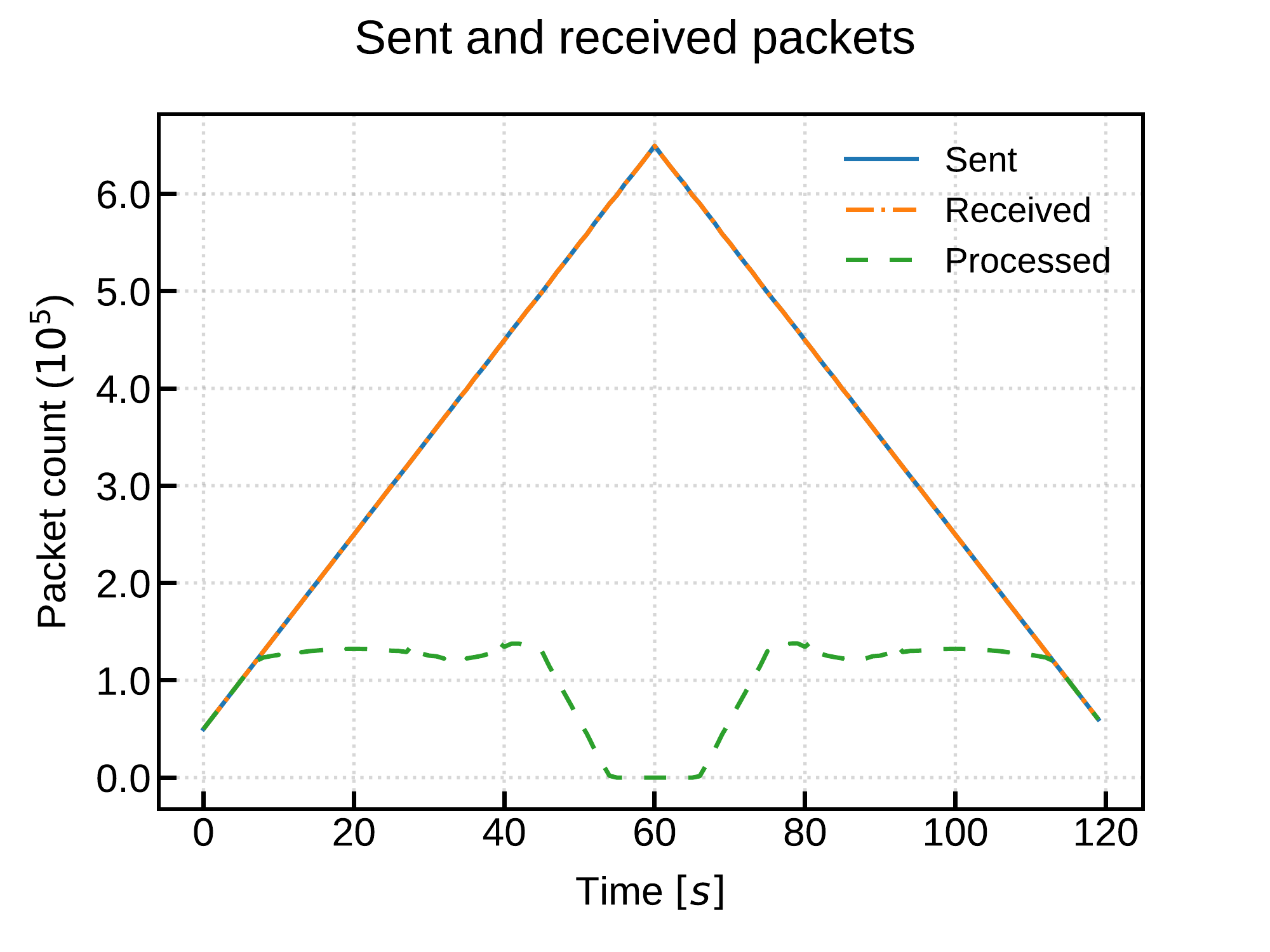}
    \caption{Scheduler: Number of sent, received, and processed packets (higher priority for critical task).}
    \label{fig:scheduler-packets}
    \end{subfigure}
    \hfill
    \begin{subfigure}[t]{0.48\columnwidth}
    %\includesvg[width=\textwidth]{plot_NONE_n400k350_CRITPRI_critical.svg}
    \includegraphics[width=\textwidth]{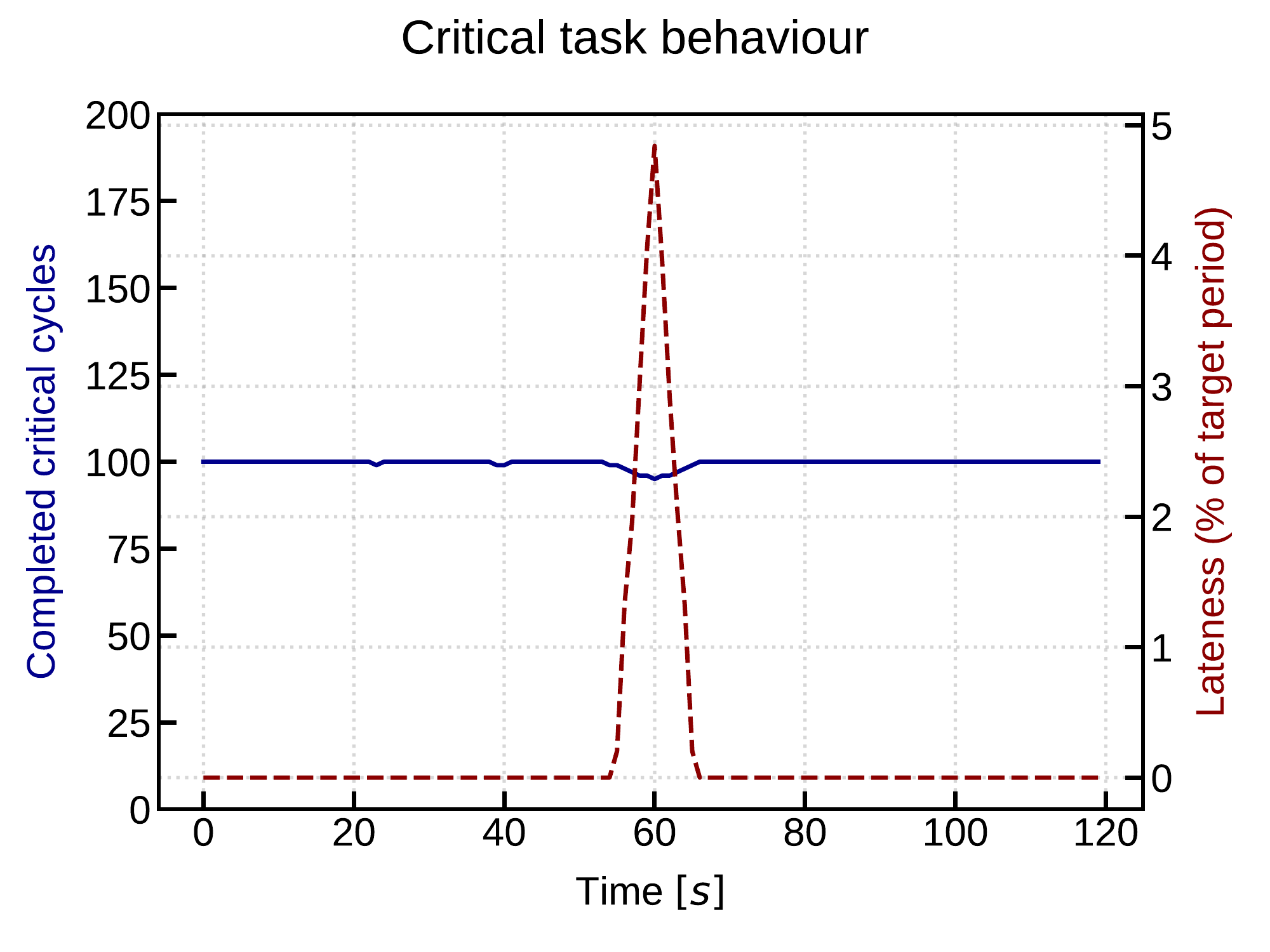}
    \caption{Scheduler: Number of completed critical cycles, accumulated lateness (same priorities as in Fig. \ref{fig:scheduler-packets}).}
    \label{fig:scheduler-critical}
    \end{subfigure}
    \\
    \begin{subfigure}[t]{0.48\columnwidth}
    %\includesvg[width=\textwidth]{plot_BRST_n400k350_CRITPRI_packets.svg}
    \includegraphics[width=\textwidth]{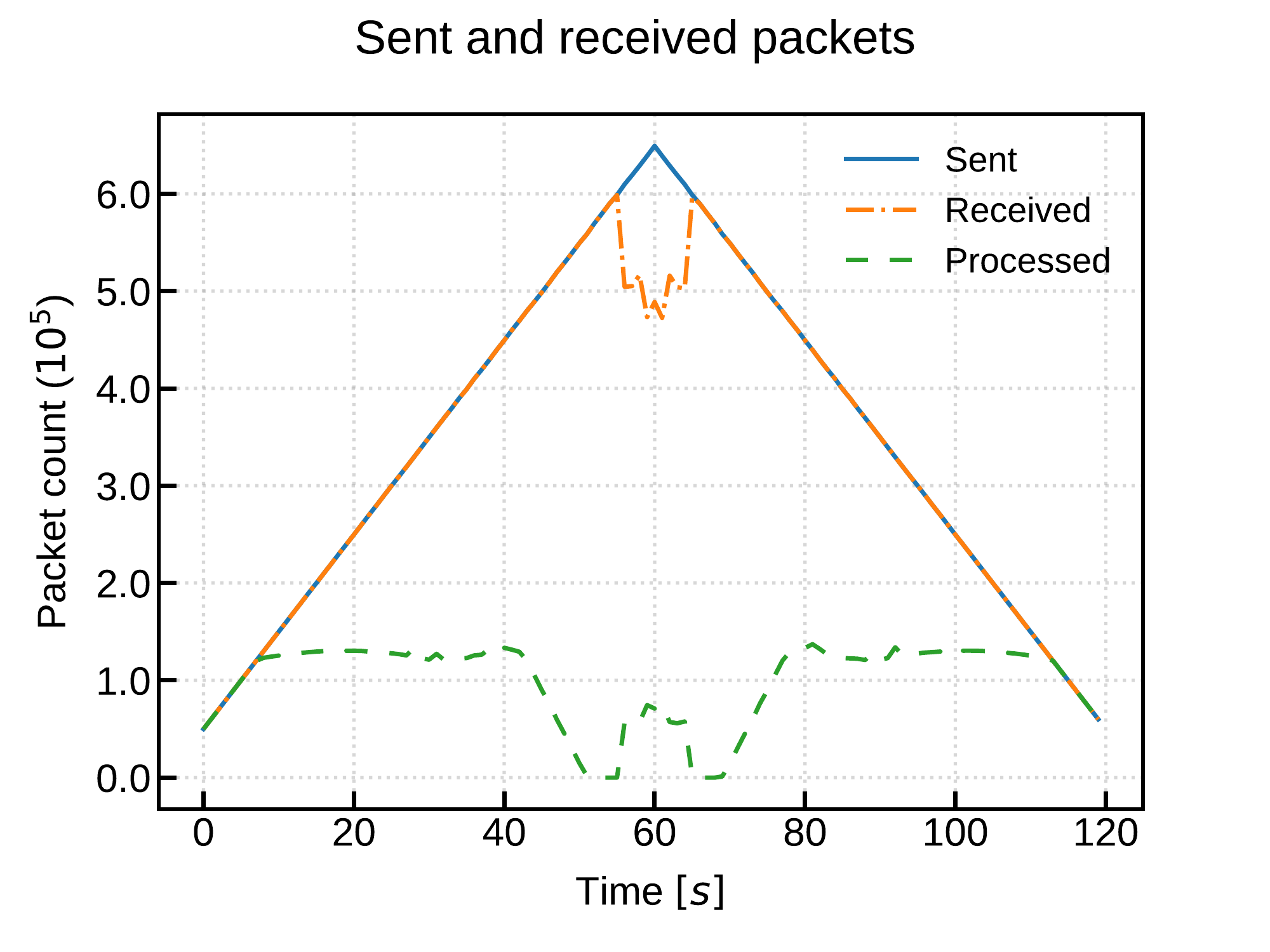}
    \caption{Burst Mitigation: Number of sent, received, and processed packets (capacity of $600$ packets per $20ms$, higher priority for critical task, queue size $100$).}
    \label{fig:burst-packets}
    \end{subfigure}
    \hfill
    \begin{subfigure}[t]{0.48\columnwidth}
    %\includesvg[width=\textwidth]{plot_BRST_n400k350_CRITPRI_critical.svg}
    \includegraphics[width=\textwidth]{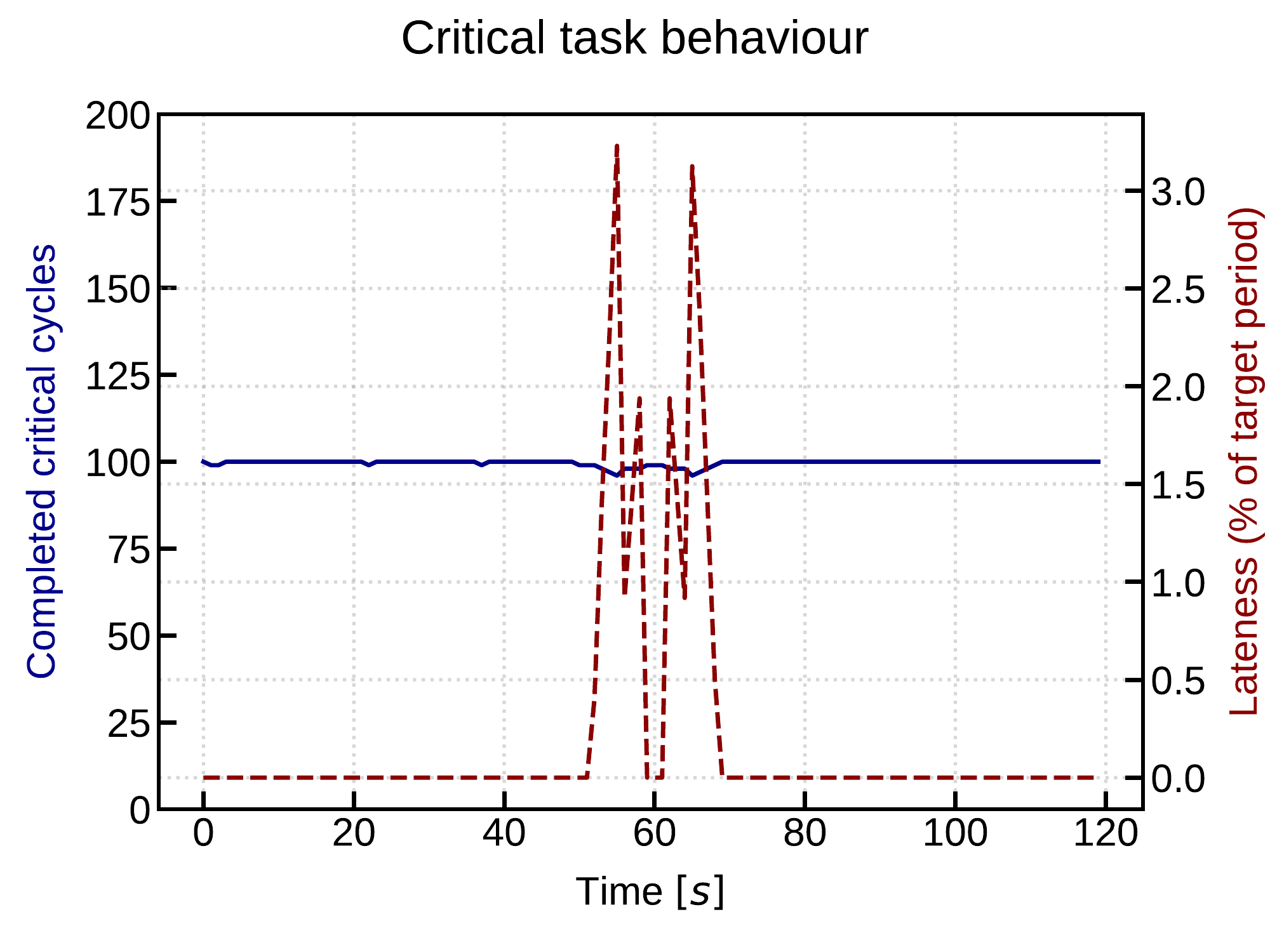}
    \caption{Burst Mitigation: Number of completed critical cycles, accumulated lateness (same parameters as in Fig. \ref{fig:burst-packets}).}
    \label{fig:burst-critical}
    \end{subfigure}
    \caption{Performance of the scheduler and Burst Mitigation}
\end{figure}

\subsubsection*{Scheduler}
Before diving into custom mitigation techniques, we take a look at the built-in FreeRTOS-Scheduler as a baseline. The option to assign (different) priorities to the critical task and network driver is a first, simple way to balance them out.

As a baseline, we tested how the system performs when both tasks are assigned equal priority.
As expected, the critical task started to incur significant lateness (up to $500\%$), while the network driver was able to process almost all packets.
Next, we tested a configuration where the critical task has higher priority than the network driver.
Interestingly, this alone brought the lateness down to almost zero.

A slight lateness of about $8\%$ is observed when the number of packets per second exceeds $50,000$, cf. Fig. \ref{fig:scheduler-critical}).
At the same time, the number of packets processed by the driver is far lower than previously, dropping even more as the interrupt count rises (cf. Fig. \ref{fig:scheduler-packets}).

\subsubsection*{Burst Mitigation}
We evaluated Burst Mitigation with a capacity of $600$ packets per $20ms$. 
We assigned a higher scheduler priority to the critical task since even though we disable the network interrupts, the network driver still processes queued packets. 
By setting the priorities in favor of the critical task, we inhibit not only the ISR but also the driver from taking too much time off the critical task. 

Figure \ref{fig:burst-packets} shows the number of sent, received, and processed packets we measured with the stated parameters.
The number of received packets per second does not exceed $30,000$, as is expected from $600$ packets per $20ms$.
In this experiment, the critical task did not introduce lateness, demonstrating the effectiveness of the mitigation when the parameter value is well chosen.
In further tests, we measured that raising the packet capacity significantly leads to lateness, showing that $30,000$ is the optimal condition for our setup.

\subsubsection*{Hysteresis Mitigation}
\begin{figure}[t]
    \centering
    \begin{subfigure}[t]{0.48\columnwidth}
    %\includesvg[width=\textwidth]{plot_HYST_n400k350_CRITPRI_packets.svg}
    \includegraphics[width=\textwidth]{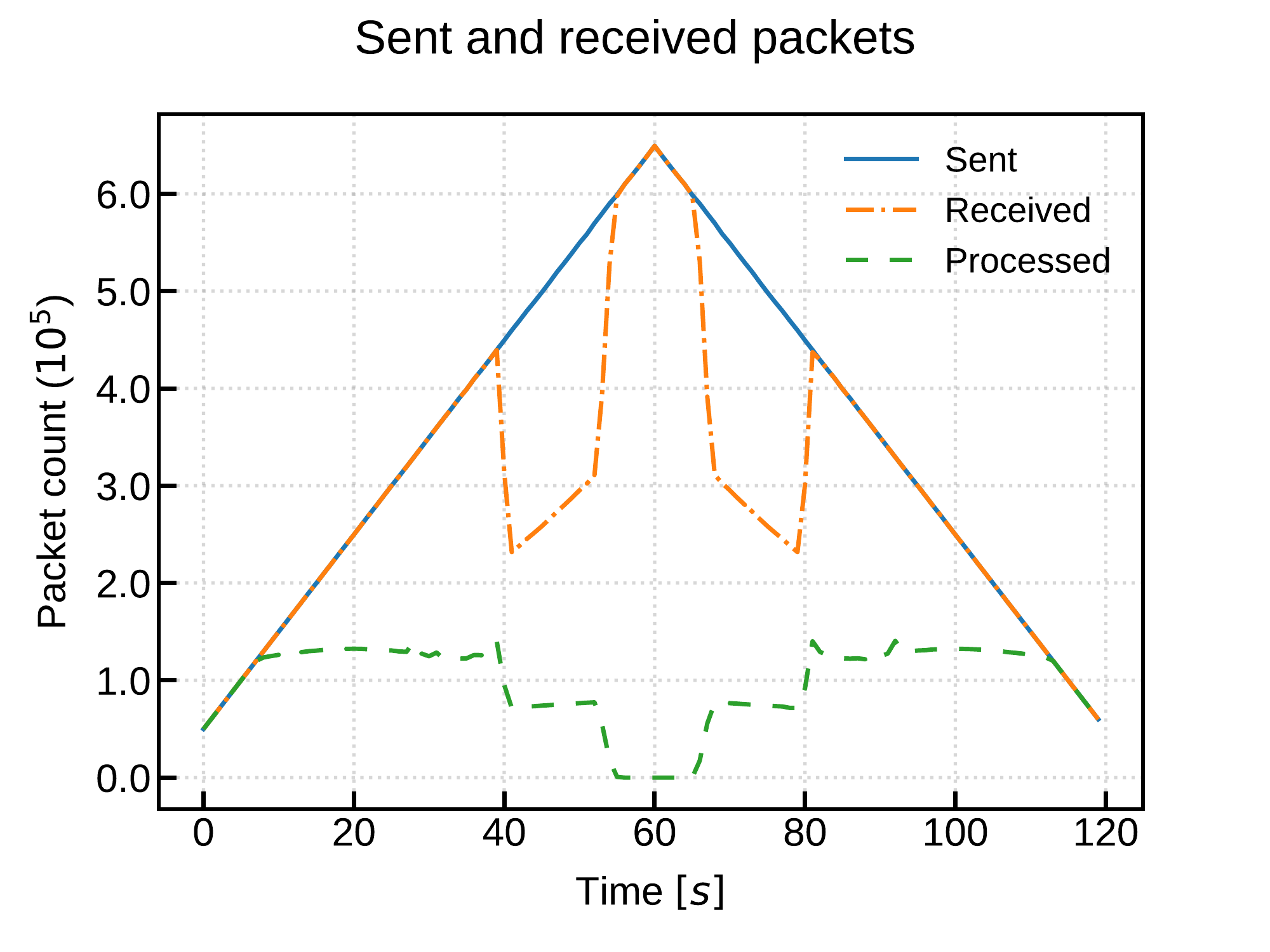}
    \caption{Hysteresis Mitigation: Number of sent, received, and processed packets (network driver deactivation at $90\%$ of allowable time expired at task completion, re-activation at $50\%$, higher priority for critical task, queue size $500$).}
    \label{fig:hyst-packets}
    \end{subfigure}
    \hfill
    \begin{subfigure}[t]{0.48\columnwidth}
    %\includesvg[width=\textwidth]{plot_HYST_n400k350_CRITPRI_critical.svg}
    \includegraphics[width=\textwidth]{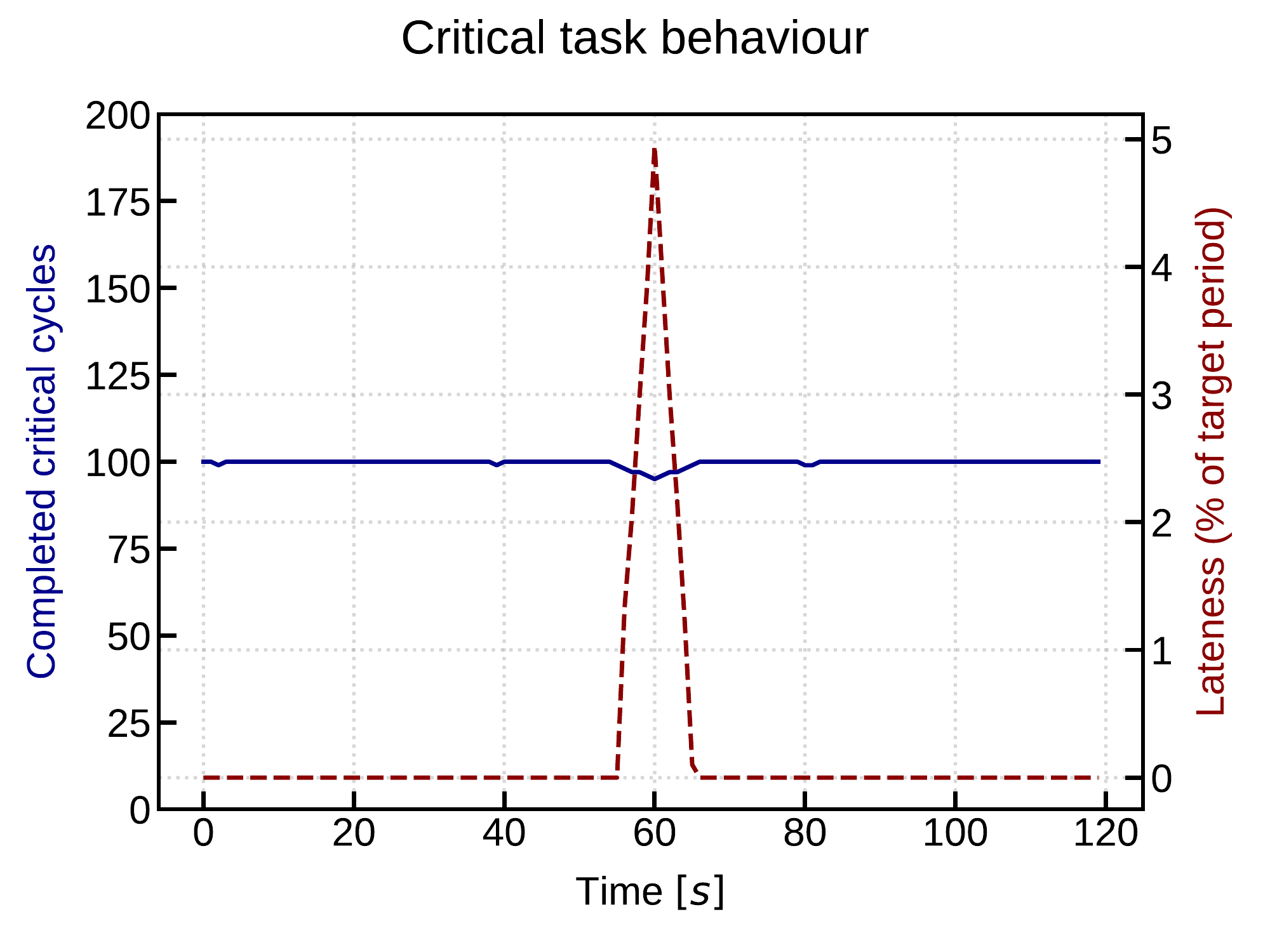}
    \caption{Hysteresis Mitigation: Number of completed cycles, accumulated lateness (same parameters as in Fig. \ref{fig:hyst-packets}).}
    \label{fig:hyst-critical}
    \end{subfigure}
    \\
    \begin{subfigure}[t]{0.48\columnwidth}
    %\includesvg[width=\textwidth]{plot_BDGT_n400k350_SAMEPRI_packets.svg}
    \includegraphics[width=\textwidth]{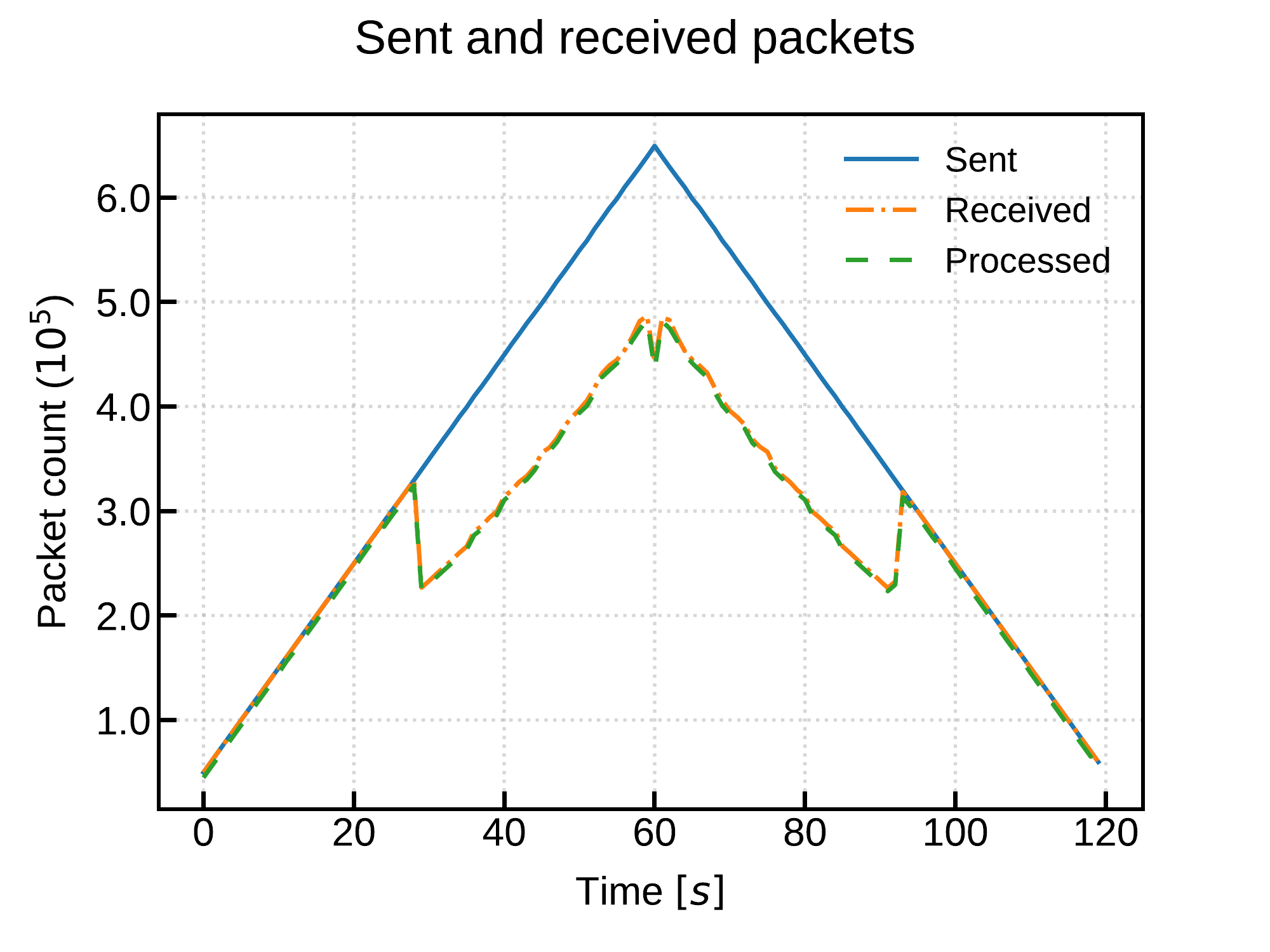}
    \caption{Budget Mitigation: Number of sent, received, and processed packets (both tasks have equal priority, queue size $100$).}
    \label{fig:budget-packets}
    \end{subfigure}
    \hfill
    \begin{subfigure}[t]{0.48\columnwidth}
    %\includesvg[width=\textwidth]{plot_BDGT_n400k350_SAMEPRI_critical.svg}
    \includegraphics[width=\textwidth]{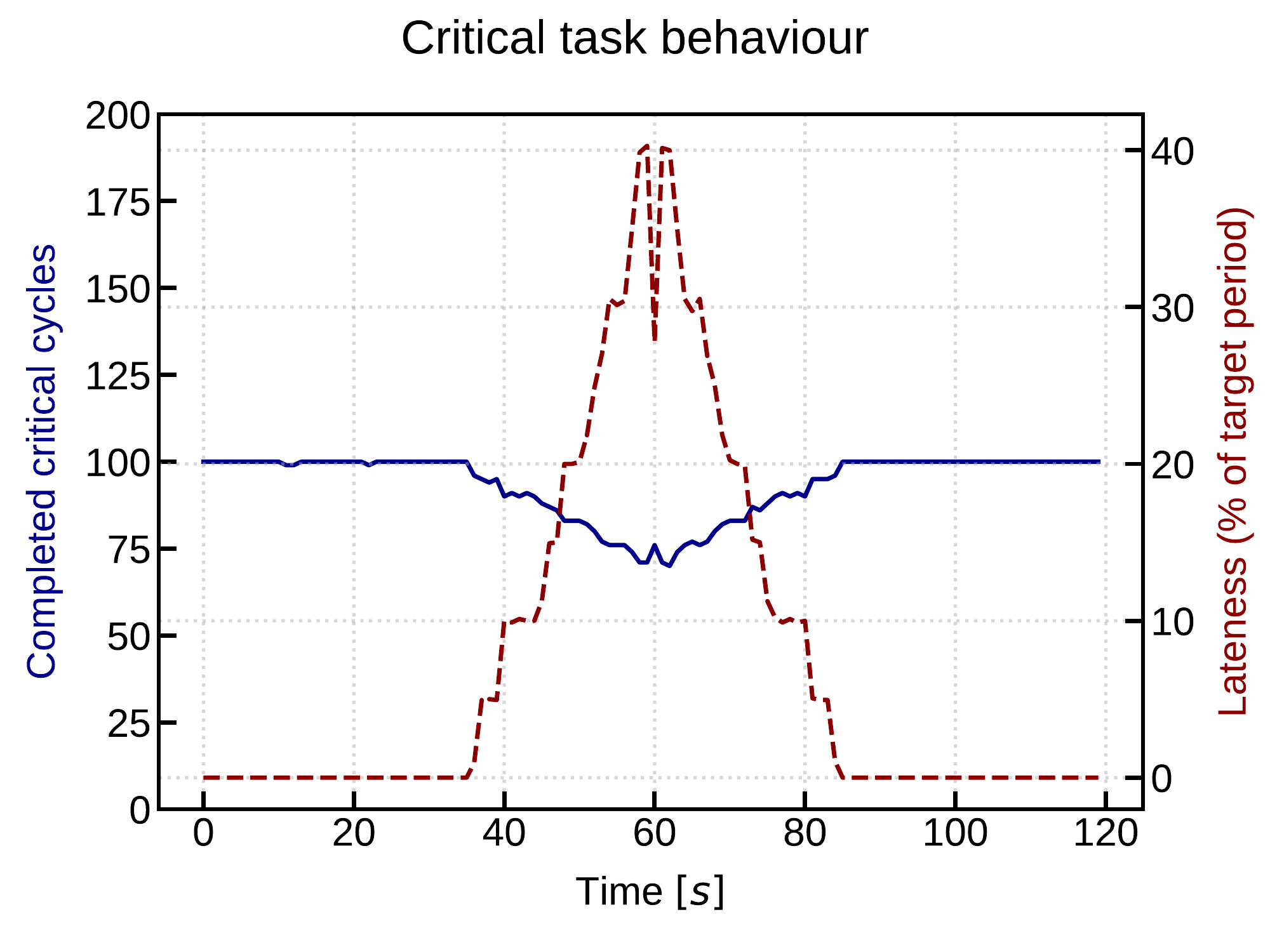}
    \caption{Budget Mitigation: Number of completed critical cycles, accumulated lateness (same parameters as in Fig. \ref{fig:budget-packets}).}
    \label{fig:budget-critical}
    \end{subfigure}
    \caption{Performance of Hysteresis and Budget Mitigation}
\end{figure}

The network driver's throughput does not stabilize as it does with the Burst Mitigation but instead plummets as the increased interrupt count puts sustained load on the CPU core Figure (cf. \ref{fig:hyst-packets}).
This does not meet expectations as interrupts should be turned off together with the processing in the network driver by the mitigation algorithm.
It turns out that the interrupts, once reactivated, will drown out the network driver where the thresholds are checked.
Once the network driver is permanently preempted by ISR executions, the mitigation algorithm can no longer take into effect, explaining why the interrupt count curve starts to fit the sent packet curve again.
Nevertheless, Figure \ref{fig:hyst-critical} shows that the Hysteresis Mitigation is quite effective in preventing lateness except when the load of interrupts itself throttles network driver execution.

\subsubsection*{Budget Mitigation} \label{sec:results-budget}
Figure \ref{fig:budget-packets} shows that once the budget is depleted for the first time, there is a drop in received and processed packets,  beyond which both curves continue to track the trend of sent packets.
The slope of the received/processed curves is less steep than that of the sent packets, suggesting that the budget has a moderating impact but cannot lower the network subsystem activity enough for the real-time guarantees to be maintained.
For each additional packet received, $0.76$ additional packets are processed.

Figure \ref{fig:budget-critical} shows that the overestimation of the budget per additional incoming packet leads to eventually breaking the real-time guarantees for high loads.

\subsubsection*{Queue Mitigation}
\begin{figure}
    \centering
    %\includesvg[width=.95\columnwidth]{plot_Q500_n400k350_CRITPRI_packets.svg}
    \includegraphics[width=.95\columnwidth]{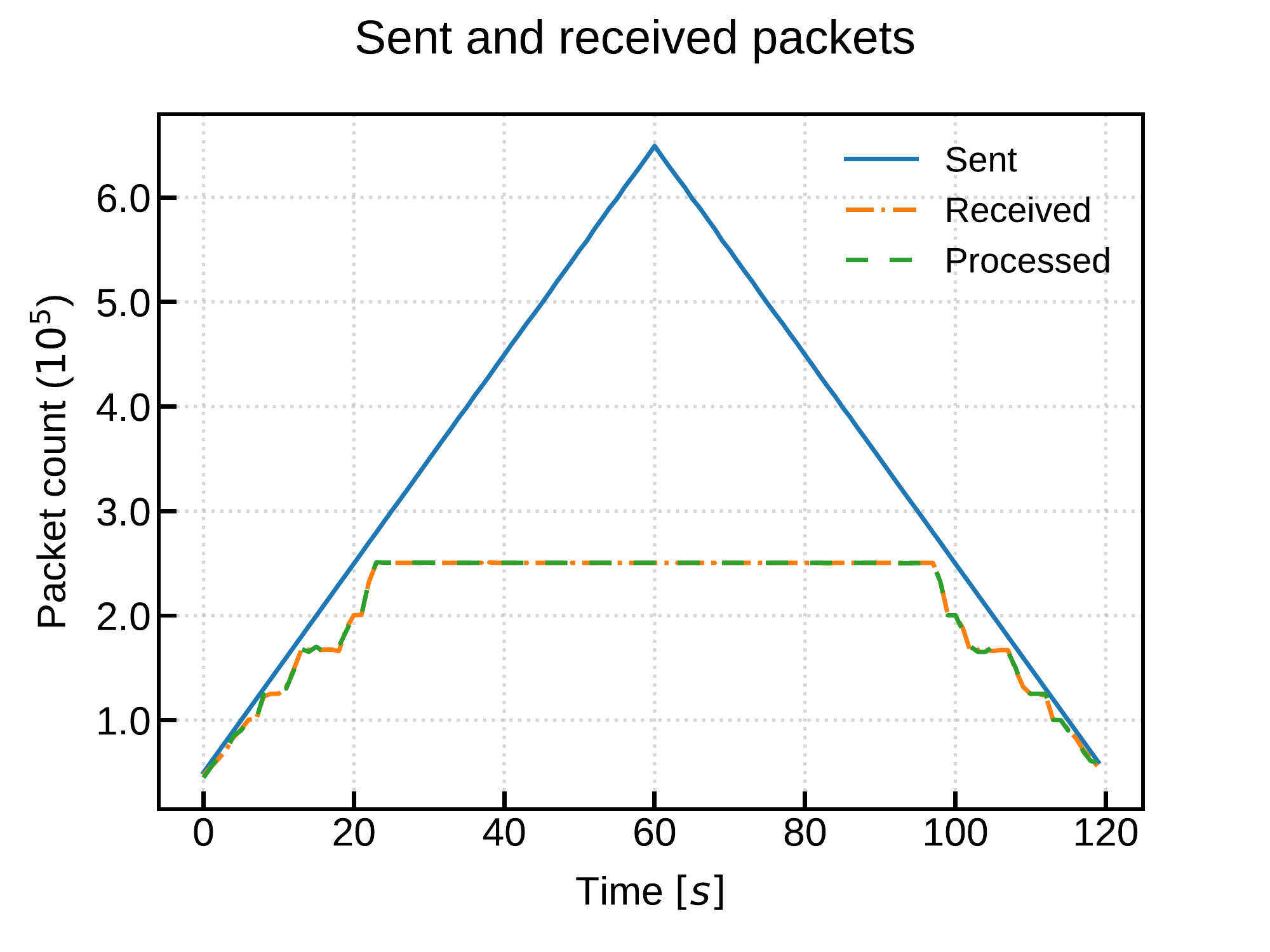}
    \caption{Queue Mitigation: Number of sent, received, and processed packets (higher priority for critical task, queue size $500$). The lateness remained at zero.}
    \label{fig:queue-packets}
\end{figure}

We evaluated the Queue Mitigation with different queue sizes.
Figure \ref{fig:queue-packets} depicts the packet numbers with a queue size of $500$.
While the network stack is not able to cope with all incoming packets once they rise above $30,000$ per second, the processed packets nicely trace the received packets.

At the same time, the critical task incurred no lateness in this setup.
Queue Mitigation was thus able to process far more packets than Burst and Hysteresis Mitigation while protecting the critical task more effectively.

We also tested smaller and larger queue sizes. Reducing the queue size to $100$ diminished packet throughput, but added no lateness.
This is because a smaller queue makes the mitigation more cautious as the queue fills up faster.
We observed the opposite effect with a queue size of $750$.
Lateness started to emerge because the mitigation reacted belatedly.

\section{Discussion} \label{sec:evaluation}

Using only the scheduler, no priority configuration led to particularly good results---with either lateness or very little packet throughput.
Additionally, its priorities cannot mitigate the problem of \emph{interrupts} drowning the process, since ISRs always run above the process priority space.
However, the scheduler priorities can still aid actual mitigation techniques in balancing out the critical task and network driver, as we will show later.
Burst Mitigation allows to avoid drowning all other computation in interrupts, but depending on how the threshold was chosen, much time is spent in the ISR receiving packets that cannot be processed in time.

Hysteresis Mitigation can be effective in environments without network interrupt loads able to drown the network driver task.
In a high-load environment, it has to be coupled with an interrupt-reducing mechanism like Burst Mitigation in the ISR, which removes some of the advantages of stand-alone usage tested above.

Unlike the Burst Mitigation, the mechanism does not require any knowledge about system throughput -- the required threshold constants generalize better across platforms with differences in processing power and are more closely aligned with the goal of minimizing lateness.

The Budget Mitigation, not unlike the FreeRTOS scheduler, accumulates more lateness the higher the load, the link is approximately linear.
This is due to its underestimation of packet processing time mentioned in Sec. \ref{sec:results-budget}.
The behavior can partially be explained by the fact that only the activity in the driver and not in the ISR is timed due to implementation difficulties.
The ISR activity thus does not impact budget consumption.

The characteristics of earliness as a reporting mechanism are also crucial: Our critical task reports the difference of the timestamps between cycle termination and cycle target, \emph{not} the time spent on the critical task alone.
If, therefore, the network driver has been allocated a large budget in one cycle, it will compete with the critical task for processing time, leading to it closely matching its deadline, and reporting little earliness.
The critical task can then terminate early the next cycle because the network task had a small budget, introducing oscillations in the budget ceiling.

Queue Mitigation always keeps the critical task on time while the packet throughput remains at a constant level.
The number of packets received by the ISR and processed by the driver is almost equal.
This indicates that interrupt handling and processing in the driver are well-balanced.

Alas, with Queue Mitigation there is still a hyper-parameter we need to tune, as with Burst Mitigation.
However, we argue that the queue size is a parameter that must be defined in any case, with any mitigation, and as argued in Section \ref{sec:queue-fill-metric} defining a fixed queue length still leaves the system more flexible and elastic than defining a fixed burst capacity.

\section{Related Work} \label{sec:related-work}

The authors of \cite{regehrPreventingInterruptOverload2005} present rate-limiting schedulers (both in software and hardware) and a burst scheduler, which is very similar to our Burst Mitigation. However, in comparison to our mitigation techniques, their approach is not specifically tuned to network interrupts and as such cannot use metrics like the queue fill state.

Many approaches to solving the problem of high interrupt counts breaking real-time priorities include extending the system with additional hardware. The Peripheral Control Processor is a proposed co-processor that executes interrupts and remaps priorities to unify the priority space between tasks and interrupts \cite{coprocessor}.

FPGA hardware monitoring and controlling the interrupt's execution of a connected microcontroller \cite{fpgacheater} can be a viable strategy. Although the proposed soft- and hardware interrupt limiters perform very well, the downside is the increased system cost.

To analyse real-time behaviour under different network loads and hardware configurations \cite{bender2021pieres} presents a playground for network interrupt experiments in IoT environments. The tool allows to run experiments on real-time embedded systems with different network interface controller implementations, load generators and timing utilities.

The approach of \cite{closebutnotcloseenough} simulates control and background traffic to a hypothetical critical power plant control system that is exposed to the internet and as such a target of DoS attacks. 
The paper presents a interrupt-overload detection. However, its mitigation techniques have the same limitations as our Burst and Hysteresis mitigation. The high processing power requirements are addressed by putting a more powerful router between the network and the embedded devices. 

The router strategy is extended by \cite{waniDDoSDetectionAlleviation2020}. The paper demonstrates how to use a software-defined network architecture for edge DDoS protection, possibly leveraging infrastructure that is already in place.

Detection can be refined by using LSTMs and CNNs for traffic classification \cite{jia2020flowguard}. Here, packets are passed through the closest edge server for execution of the models.

\section{Conclusion}\label{sec:conclusion}
The trend of putting embedded devices at the edge of the network to perform critical tasks can expose the whole system to risk. The edge devices may fail to hold up to a sudden surge in traffic or be targeted by denial of service attacks.

We analyzed the network interrupt count and queue fill state as well as the lateness of a critical task running on an embedded SoC while it was flooded with network traffic. We developed four mitigation strategies to maintain smooth operation and timeliness of a critical process using signals derived from the data obtained during the analysis: Burst Mitigation, Hysteresis Mitigation, Budget Mitigation, and finally, Queue Mitigation.

In our experiments, we measured the quality of the different mitigation strategies through the lateness of a critical task that runs concurrently to a network simulation triggering a high amount of interrupts. An evaluation has shown that Queue Mitigation performed the best since it receives and processes the most network packets while also protecting the critical task from lateness.

\bibliographystyle{ACM-Reference-Format}
\bibliography{refs}

\end{document}